\documentclass[prl,twocolumn,a4paper,showpacs,preprintnumbers]{revtex4}
\usepackage{citesort}
\usepackage{epsfig}
\usepackage{amsmath,amssymb}
\newcommand{\nn}{\nonumber}
\newcommand{\dsl}[1]{\hspace{#1}/}
\allowdisplaybreaks
\begin{document}
\preprint{KEK-TH-1081}
\preprint{TU-766}
\preprint{hep-ph/0604208}
\title{Parity-Odd Asymmetries in $W$-Jet Events at the Tevatron}
\author{Kaoru Hagiwara,$^{1}$ Ken-ichi Hikasa,$^{2}$ and Hiroshi Yokoya$^{3}$}
\affiliation{\\
$^{1}$KEK Theory Division and GUAS (Sokendai), Tsukuba 305-0801, Japan\\
$^{2}$Department of Physics, Tohoku University, Aoba-ku, Sendai
980-8578, Japan\\
$^{3}$Department of Physics, Niigata University, Niigata 950-2181, Japan
}
\date{\today}
\begin{abstract}
Parity-odd asymmetries in the decay angular distribution of a $W$ boson 
produced with a hard jet in $p\bar p$ collisions arise only from QCD 
rescattering effects.  If observed, these asymmetries will provide a first 
demonstration that perturbative QCD calculation is valid for the absorptive 
part of scattering amplitudes.  We propose a simple observable to 
measure these asymmetries and perform realistic Monte Carlo simulations 
at Tevatron energies.  It is shown that the Tevatron Run-II should 
provide sufficient statistics to test the prediction.
\end{abstract}
%
\maketitle
%
Weak boson production at hadron colliders has become a high statistics 
process which allows for detailed studies of the underlying dynamics.  
A substantial fraction of the weak bosons have large 
transverse momentum (high-$q_T$) with recoiling jets~\cite{ua1,ua2,cdf0,d0}.  
The data are reproduced well by QCD calculations~\cite{arnold1,gonsalves}
based on the factorization of long- and short-distance physics. 
Recently, azimuthal angle distributions of charged leptons arising 
from these $W$ bosons were measured by the CDF Collaboration~\cite{cdf} 
using the Run I data at the Tevatron $p\bar{p}$ collider. 
The results are in good agreement with the predictions~\cite{strologas} 
including higher order corrections 
up to ${\cal O}(\alpha_s^2)$~\cite{mirkes1}. 
The decay polar angle distributions have been also measured by 
D0~\cite{d001} and CDF~\cite{cdf04} Collaborations.  
These lepton distributions can provide extra information 
on the production mechanism, because the angular distributions of the 
decay leptons reflect the polarization of the produced weak boson, and 
especially the interference between different polarization states.  

Parity-nonconserving ($P$-odd) asymmetries in the lepton distributions 
merit particular attention, even though the Run I luminosity was too
small to measure them.
As is well known, $P$-odd asymmetries without a spin measurement 
are also na\"\i ve-$T$-odd observables, because both $P$ and $T$
transformations change the sign of any three-momentum~\cite{naive}.
In $T$-conserving theories such asymmetries can arise from the
scattering phase, or the absorptive part of the amplitudes~\cite{hhk2}.
Although one may argue that perturbative QCD calculations of the absorptive 
part of scattering amplitudes were validated in deep inelastic scattering 
and $e^+e^-$ annihilation processes, it is limited to the {\it forward\/} 
scattering amplitudes and no experimental test has been made for the 
absorptive part of {\it non-forward\/} amplitudes.  
Measurements of such $P$-odd quantities thus would test 
perturbative QCD in a new regime and give support to the calculations 
of strong interaction phases e.g.\ in $B$ decays, where the interplay of 
strong and weak phases produces $CP$ violating asymmetries.  

The one-loop absorptive part of the high-$q_T$ $W$ plus jet amplitudes were 
calculated in Ref.~\cite{hhk1} more than two decades ago.  
The $P$-odd asymmetries in the decay lepton angular distributions were found 
to arise at ${\cal O}(\alpha_s)$ and their $q_T$ dependence was presented.  
In realistic experiments, however, 
the asymmetries calculated in Ref.~\cite{hhk1} suffer from a cancellation 
problem arising from the ambiguity in kinematic reconstruction of the 
$W$ momentum with unknown neutrino longitudinal momentum.  
In this paper, we propose new observables corresponding to the $P$-odd 
asymmetries which can be measured without kinematical ambiguity, 
and perform realistic Monte Carlo simulation 
including the effects of QCD higher-order corrections and appropriate 
kinematical cuts for detector acceptance and event selection. 
The predictions made in Ref.~\cite{hhk1} were for the Sp$\bar{\rm p}$S 
collider at the center-of-mass (c.m.) energy of 540 GeV.  
We here reevaluate the asymmetries at the Tevatron energy of 1.96 TeV.

We consider production of a high-$q_T$ $W$ boson associated by a recoiling 
jet in $p\bar p$ collisions: 
\begin{align}
& p\ (p_p) + \bar{p}\ (p_{\bar{p}})\,\to\,W^-\ (q) + {\rm jet}\ (p_j) + X;
 \\[2mm] 
& W^-\ (q) \to \ell^{-}\ (p_{\ell}) + \bar{\nu}\ (p_{\bar{\nu}})\ . \nn 
\end{align}
The spin-averaged differential cross section can be cast in the following form 
in which the dependence on $W$ decay angles are made explicit:
\begin{align}
\label{MasterDistribution}
 \frac{d\sigma}{dq_T^2d\cos\hat\theta d\cos\theta d\phi} &=
 F_1(1+\cos^2\theta) + F_2(1-3\cos^2\theta) \nn\\
 &+ \ F_3\sin{2\theta}\cos\phi + F_4\sin^2\theta\cos{2\phi} \nn\\[2mm]
 &+ \ F_5\cos\theta + F_6\sin\theta\cos\phi \nn\\[2mm]
 &+ \ F_7\sin\theta\sin\phi + F_8\sin{2\theta}\sin\phi \nn\\[2mm]
 &+ \ F_9\sin^2\theta\sin{2\phi} .
\end{align}
There are nine independent functions reflecting the spin-1 nature of 
the $W$.  
In Eq.~(\ref{MasterDistribution}), 
$q_T$ and $\hat{\theta}$ are the transverse momentum and the 
 scattering angle of the $W$ boson in the $W$-jet c.m.\ frame.  
The polar and azimuthal angles $\theta$ and $\phi$ of the charged 
lepton is defined in the Collins-Soper frame~\cite{cs} in this paper. 
Collins-Soper frame is a rest frame of the $W$ boson in which the $z$-axis 
is taken to bisect the opening angle between $\vec{p}_{p}$ and 
$-\vec{p}_{\bar{p}}$, and the $y$-axis is along the direction of
 $\vec{p}_p\times(-\vec{p}_{\bar{p}})$~\cite{csf}.  
The azimuthal angle is measured from the $x$-axis which lies in the 
 scattering plane. 

The $F_1$ term gives the total rate of the high-$q_T$ $W$ production 
after integration over the lepton angles: 
${d\sigma}/{dq_T^2d\cos\hat\theta}={16\pi}/{3}F_1$. 
The terms $F_1$ through $F_6$ are $P$-even, and the rest are $P$-odd. 
The nine invariant functions $F_{i}$ are written 
 as a convolution of the parton distribution functions and the hard scattering 
part:
\begin{equation}
 F_{i} = \sum_{a,b}\int dY f_{a/p}(x_+,\mu_F^2)\ f_{b/\bar{p}}(x_-,\mu_F^2)\ 
 \hat{F}_{i}^{ab\to W^{-}j} ,
\end{equation}
 where the momentum fractions $x_{\pm}$ are 
\begin{equation}
 x_{\pm} = 
 \frac{q_T+(q_T^2+m_W^2\sin^2\hat\theta)^{1/2}}
 {s^{1/2}\sin\hat\theta}\ e^{\pm Y}.
\end{equation}
The scale of the distribution functions $\mu_F$, as well as the scale of
 the running strong coupling constant $\mu_R$, should be
 set to the relevant scale of the collisions. 
The hard part functions $\hat{F}_i$ are expressed as
\begin{equation}
\hat{F}^{ab\to W^-j}_{i} = \frac{3\,B\,G_F\,m_W^2}
{4\sqrt{2}s(\hat{s}+m_W^2)\sin^2\hat\theta}\ f^{ab\to W^-j}_{i}(x_a,x_b)
\end{equation}
where $\hat{s}=x_+x_-s$, $B=B(W^-\to \ell^-\bar\nu_\ell)$, and 
$G_F$ the Fermi constant.  
$f^{ab\to W^{-}j}_i$ are the functions of the dimensionless variables
 $x_a=m_W^2/2q\cdot p_a$ and $x_b=m_W^2/2q\cdot p_b$.  
$p_a$ ($p_b$) is momentum of a parton $a$ ($b$) inside the 
proton (antiproton).  

In the leading order, the annihilation subprocess 
 $q\bar{q}' \to W^- g$ and the Compton subprocess 
 $qg \to W^- q'$ ($\bar{q}g \to W^- \bar{q}'$) contribute. 
For both processes, the leading contribution to $i=1$--6 comes from the tree
 level diagrams~\cite{chaichian}, and the remainings ($i=7,8,9$) 
 receive leading contribution from the one-loop diagrams~\cite{hhk1}. 
In our notation, $f_i$ with $i=1$ to 6 are ${\cal O}(\alpha_s)$, while
 $f_{7,8,9}$ are ${\cal O}(\alpha^2_s)$, at the leading order. 
Complete analytic forms in our notation can be found  e.g.\ in 
Refs.~\cite{hhk1} and \cite{hky}, where in the latter reference 
the $Z$ boson vertices are found.  

Higher order corrections are important for quantitative predictions.  
In particular, recoil logs dominate the higher order corrections 
at small $q_T$~\cite{arnold2,ellis1,ellis2}, so that all-order resummation 
is needed to gain a reliable prediction~\cite{balas,kulesza1,kulesza2}.
For the total rate of $W$-jet events, the 
 NLO correction is known to give an enhancement of 
$K \sim 1.3$~\cite{arnold1,gonsalves} relatively independent of 
$q_T$ at $q_T>30$ GeV. 
The higher order effects are known to be well approximated by setting
$\mu_F=\mu_R=q_T/2$ in the LO expression~\cite{bawa}.  
Moreover, threshold resummation studies indicate rather modest NNLO 
corrections~\cite{kidonakis}.  
The NLO correction to the $P$-even parts of the angular distribution are 
found to be small~\cite{mirkes1,mirkes2}, while those to the $P$-odd 
parts have not been calculated to our knowledge.  
Later we will limit ourselves to $W$ bosons with $q_T>30$ GeV because 
of large higher order corrections for $q_T\alt 20$ GeV~\cite{boer}. 

\begin{figure}[t]
\vspace{10pt}
\epsfig{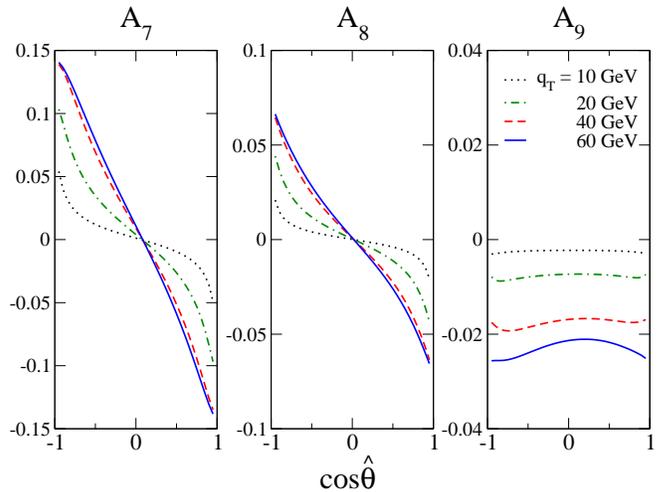}
\caption{Differential asymmetries $A_7$ (left), $A_8$ (center), and
 $A_9$ (right) for $W^-$ production in $p\bar{p}$ collision at
 $\sqrt{s}=1.96$ TeV.
$\hat\theta$ denotes the scattering angle between proton beam and $W^-$
 momenta in the $W$-jet c.m. frame. }\label{fig1} 
\end{figure}

In Fig.~\ref{fig1}, we show the differential $P$-odd asymmetries,
 defined as 
\begin{equation}
 A_{i}(q_T,\cos\hat\theta)=F_{i}/F_1\label{asy}
\end{equation}
 with $i=7,8,9$.  
This is the Tevatron Run-II adaptation of Fig.~2 of Ref.~\cite{hhk1},
 with CTEQ6M parton distribution functions~\cite{cteq}.  
The asymmetries grow monotonically with increasing $q_T$.  
The curves of $A_{7,8}$ ($A_9$) are nearly anti-symmetric 
 (symmetric) under the reflection of the sign of $\cos\hat\theta$.
Interestingly, the individual asymmetries (not shown) of
 the annihilation subprocess and the Compton subprocess are very similar. 
The largest asymmetry is expected for $A_7$ where the magnitude of the
 asymmetry exceeds 10\% at $q_T=40$ GeV for larger values of
 $|\cos\hat\theta|$. 
The asymmetries for the $W^{+}$ production are obtained from those of
 $W^{-}$ by $CP$ transformation ($\hat\theta \to \pi - \hat\theta$,
 $\theta \to \pi - \theta$, $\phi \to \phi$). 

As we cannot measure the neutrino longitudinal momentum at
 hadron colliders, there is a two-fold ambiguity on the sign of
 $\cos\theta$ in determining the Collins-Soper frame. 
This ambiguity affects $A_8$, but 
$A_7$ and $A_9$ are independent of the sign of $\cos\theta$. 
However, we have another two-fold ambiguity in determining $\cos\hat\theta$ 
and the asymmetries of Eq.~(\ref{asy}) cannot be measured directly. 
Instead of $\cos\hat\theta$, we propose to use the difference of the
 pseudorapidities of the charged lepton and the jet in the laboratory frame,
 $\Delta\eta\equiv \eta_{\ell}-\eta_{j}$, which has a positive correlation 
with $\cos\hat\theta$.  

In our analysis, we include kinematical cuts for detecting 
charged leptons and jets~\cite{strologas,mirkes2}.  
For $W \to \ell\bar{\nu}_{\ell}$ event selection, 
 we apply the following cuts: $|\eta_{\ell}^{\rm lab}|<1$, 
$E^{\ell}_{T} > 20$ GeV, $E\dsl{-6pt}_{T} > 20$ GeV and 
$M^{W}_{T} > 40$ GeV, where $E\dsl{-6pt}_{T}$ is the missing transverse 
 energy and $M^{W}_{T}=\sqrt{(|p^{\ell}_{T}|+|p\dsl{-5pt}_T|)^2
 -(\vec{p}^{\ell}_{T}+\vec{p}\dsl{-5pt}_{T})^2}$ 
 the transverse mass of the $W$.  
For jet detection and identification, we demand; 
 $|\eta_{j}^{\rm lab}| < 2.4$, $E^{j}_{T} > 15$ GeV and $\Delta{R} > 0.7$, 
where $\Delta{R} \equiv  \sqrt{(\Delta\eta)^2 +(\Delta\phi)^2}$, with 
$\Delta\phi$ the difference between the azimuthal angles of the charged 
lepton and the jet in the laboratory frame~\cite{cdf}. 

We perform Monte Carlo simulation using BASES/SPRING~\cite{bases} with 
an integrated luminosity of 1 fb$^{-1}$.  
We consider a single flavor of leptonic $W$ decay (either $e$ or
$\mu$), and combine $W^+$ and $W^-$ events by assuming $CP$
invariance~\cite{cp}.  
The event yields are found to be $(6, 10, 11, 8) \times 10^3$ 
for $30<q_T ({\rm GeV}) <50$ and 
$(2, 5, 5, 3) \times 10^3$ for $q_T ({\rm GeV}) > 50$, respectively 
in the $\Delta\eta$ intervals 
(${<}{-1}$, [$-1$, 0], [0, 1], ${>}1$).  
At the Tevatron energy, the contributions of annihilation
 subprocess and Compton subprocess to the production cross section is  
 about the same~\cite{arnold1,gonsalves}. 

We examine two observables related to the $P$-odd asymmetries, which can 
be defined without knowing the neutrino longitudinal momentum.  
One is a left-right asymmetry of the charged lepton momentum with
 respect to the scattering plane; 
\begin{equation}
 A_{LR}(\Delta{\eta}, q_T) \equiv
  \left[N(\vec p_{\ell}\cdot\vec n_y >0) 
- N(\vec p_{\ell}\cdot\vec n_y <0)\right]/\ N_{\rm sum},\label{eqalr} 
\end{equation}
where $\vec n_y = \vec p_p {\times} \vec q_T
 /|\vec p_p {\times} \vec q_T|$ in the laboratory frame.  
It is defined as the asymmetry between the number of events 
having charged lepton momentum with positive and negative $y$ component, 
and may be expressed as 
$[N(0<\phi<\pi)-N(\pi<\phi<2\pi)]/N_{\rm sum}$ in the Collins-Soper frame.  
Since $\sin\theta\sin\phi$ is proportional to
 the $y$-component of the charged lepton momentum,
 this asymmetry is expected to reflect the property of $A_7$. 
\begin{figure}[t]
\vspace{20pt}
\begin{center}
 \epsfig{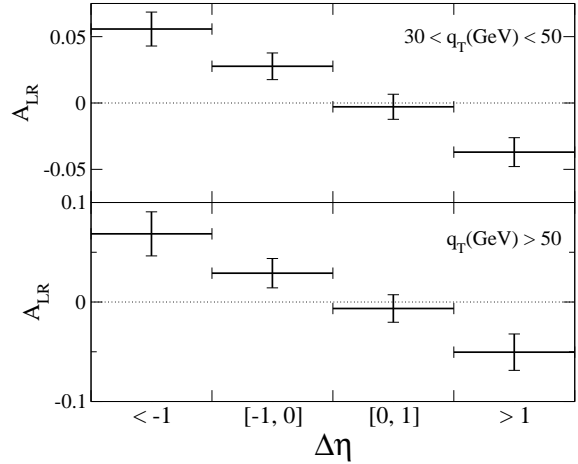}
\caption{Left-right asymmetries defined in Eq.~(\ref{eqalr}) for
 $30<q_T ({\rm GeV}) < 50$ (upper), and for $q_T ({\rm GeV})>50$ (lower).
$\Delta{\eta}=\eta_{\ell}-\eta_j$ is the difference of the $\ell^-$ and jet
 pseudorapidity in the laboratory frame.
In the figure, asymmetries are evaluated by setting the scale as
 $\mu=q_{T}/2$, and error-bars show the expected statistical errors for
 the 1 fb$^{-1}$ luminosity at the Tevatron.}\label{fig2}   
\end{center}
\end{figure}
In Fig.~\ref{fig2}, we show the left-right asymmetries with expected
 statistical error-bars obtained by assuming 1 fb$^{-1}$ luminosity and
 setting the scale as $\mu=q_T/2$.
The errors are estimated from $\delta{A}=\sqrt{(1-A^2)/N_{\rm sum}}$ for
 each bin.
The asymmetry is negative for $\Delta\eta>0$ and positive for
 $\Delta\eta<0$, which agrees with the shape of $A_7$ {\it vs.}\ 
 $\cos\hat\theta$ in Fig.~\ref{fig1} (left) as is expected. 
The magnitude of the asymmetry increases with $q_T$, 
 and reaches 5\% for $|\Delta\eta|>1$, which is about  
 30\% of the magnitude of $A_7$ at large $|\cos\hat\theta|$.

Since our predictions for the $P$-odd asymmetries are at the LO level,
they can have significant higher-order corrections.
We estimate their uncertainties by varying the renormalization and
factorization scale by a factor 2 and 1/2. 
The results listed in Table I, show that our predictions are uncertain
by 10--20\%.

It may be useful to estimate the minimum luminosity needed to establish
the non-zero asymmetry at the Tevatron.
To define the significance, we simply combine the eight bins of Fig.\ 2
as $\chi^2=\sum_{\rm bins}A^2/(\delta{A})^2$.
By using the asymmetry values listed in Table I, we find that
the integrated luminosity needed to establish the non-zero asymmetry at
the $3\sigma$ level ($\chi^2>9$) is 250, 180, 150 pb$^{-1}$ when we
calculate the LO expression at $\mu=q_T$, $q_T/2$, $q_T/4$, respectively.
Although the above numbers are obtained by assuming 100\% detecion
efficiency with no systematic errors, the prospect of observing the
predicted asymmetry in the Run II data should be good, since more
luminosity will be accumulated and both electron and muon decays can be
used at two detectors.\\
\begin{table}[b]
\begin{center}
\renewcommand{\arraystretch}{1.4}
\begin{tabular}{|c|*{4}{r@{\hspace{5pt}}|}}
\hline
$\Delta{\eta}=\eta_{\ell}-\eta_{j}$&
\ $<-1$&[$-1$, $0$]&[$0$, $1$]&\ $>1$\\
\hline
& 4.7& 2.2& $-0.8$& $-3.3$\\
$30 < q_T\,(\rm GeV) < 50$
& 5.3& 2.4& $-0.9$& $-3.7$\\
& 6.2& 2.9& $-1.0$& $-4.3$\\
\hline
& 5.1& 2.4& $-0.6$& $-4.4$\\
$q_T\,(\rm GeV)>50$
& 5.9& 2.5& $-0.7$& $-4.2$\\
& 6.9& 2.9& $-1.1$& $-4.5$\\
\hline
\end{tabular}
\end{center}
 \caption{Table of the scale ambiguities of the left-right asymmetries
 $A_{LR}$ $(\%)$. The asymmetries are calculated in LO with setting the
 scale as $\mu=q_T$ (upper), $q_T/2$ (middle) and $q_T/4$ (lower) for
each bins.}\label{tab1}\end{table}

Another asymmetry may be defined as
\begin{align}
A&_{Q} \equiv \left[N(0\!<\!\phi\!<\!\pi/2)
 -N(\pi/2\!<\!\phi\!<\!\pi)\right. \nn \\[2mm]
&+\left.
 N(\pi\!<\!\phi\!<\!3\pi/2)
 -N(3\pi/2\!<\!\phi\!<\!2\pi)\right] /\ N_{\rm sum}. 
\end{align}
Since $\sin2\phi$ is positive for $0\!<\!\phi\!<\!\pi/2$ and
$\pi\!<\!\phi\!<\!3\pi/2$, and negative for $\pi/2\!<\!\phi\!<\!\pi$
and $3\pi/2\!<\!\phi\!<\!2\pi$, this quadrant asymmetry reflects the
nature of $A_9$. 
$A_9$ is expected to have the same sign for any $\cos\hat\theta$, 
therefore we may combine all data regardless of $\Delta\eta$.  
For this reason and the smallness of the expected magnitude of $A_9$, we
calculate this asymmetry from all the events with any 
$\Delta\eta$ and $q_T>30$ GeV, and find 
$A_{Q}=-0.8\%$ ($\mu=q_T$), $-0.8\%$ ($\mu=q_T/2$) and $-0.9\%$
($\mu=q_T/4$). The statistical error is expected to be $\pm 0.5\%$ for
the 1 fb$^{-1}$ luminosity at the Tevatron. 

Summing up, we have studied the $P$-odd asymmetries in the decay angular
distribution of a $W$ boson produced with a hard jet in $p\bar{p}$
collisions.
We have proposed simple observables which are free from the ambiguity in
the neutrino longitudinal momentum, and performed a realistic Monte Carlo
simulation.
It was shown that these asymmetries can be measured at the Tevatron
Run-II, with the recent increase in the luminosity. 

\begin{acknowledgments}
We thank K.~Sato for useful discussions. 
H.Y.\ wishes to thank K.~Mawatari for helpful guide 
of using the BASES/SPRING code.
The first K.H.\ wishes to thank Aspen Center for Physics and the
organizers of the Aspen 2005 Collider Physics Workshop. 
The work of second K.H.\ is supported in part by the Grant-in-Aid for 
Scientific Research (No.\ 14046201) from the Japan Ministry of Education, 
Culture, Sports, Science, and Technology.
The work of H.Y.\ is supported in part by a Research Fellowship 
of the Japan Society for the Promotion of Science. 
\end{acknowledgments}


%

\begin{thebibliography}{99}
\bibitem{ua1}
C.~Albajar {\it et al.}\  [UA1 Collaboration],
Z.\ Phys.\ C {\bf 44}, 15 (1989).
%
\bibitem{ua2}
J.~Alitti {\it et al.}\  [UA2 Collaboration],
Z.\ Phys.\ C {\bf 47}, 523 (1990).
%
\bibitem{cdf0}
F.~Abe {\it et al.}\ [CDF Collaboration],
Phys.\ Rev.\ Lett.\ {\bf 66}, 2951 (1991); 
{\it ibid} {\bf 67}, 2937 (1991).
%
\bibitem{d0}
B.~Abbott {\it et al.}\ [D0 Collaboration],
Phys.\ Rev.\ Lett.\ {\bf 80}, 5498 (1998).
%
\bibitem{arnold1}
P.~B.~Arnold and M.~H.~Reno,
Nucl.\ Phys.\ B {\bf 319}, 37 (1989);\ B {\bf 330}, 284(E) (1990).
%
\bibitem{gonsalves}
R.~J.~Gonsalves, J.~Pawlowski and C.~F.~Wai,
Phys.\ Rev.\ D {\bf 40}, 2245 (1989); 
Phys.\ Lett.\ B {\bf 252}, 663 (1990).
%
\bibitem{cdf}
D.~Acosta {\it et al.}\ [CDF Collaboration], Phys.\ Rev.\ D {\bf 73}, 052002 
(2006).
%
\bibitem{strologas}
J.~Strologas, and S.~Errede, Phys.\ Rev.\ D {\bf 73}, 052001 (2006).
%
\bibitem{mirkes1}
E.~Mirkes, Nucl.\ Phys.\ B {\bf 387}, 3 (1992); 
E.~Mirkes, J.~G.~K\"{o}rner and G.~A.~Schuler,
Phys.\ Lett.\ B {\bf 259}, 151 (1991). 
%
\bibitem{d001}
B.~Abbott {\it et al.}\ [D0 Collaboration], Phys.\ Rev.\ D {\bf 63}, 
072001 (2001).
%
\bibitem{cdf04}
D.~Acosta {\it et al.}\ [CDF Collaboration], Phys.\ Rev.\ D {\bf 70}, 
032004 (2004).
%
\bibitem{naive}
What we call na\"\i ve $T$-odd asymmetry here is sometimes referred to as 
$T$-odd asymmetry.  We use the word `na\"\i ve' to remind that nonvanishing 
asymmetry does not necessarily imply $T$ violation.  This is because 
$T$ transformation interchanges initial and final states, whereas 
the asymmetry is defined without such interchange.
%
\bibitem{hhk2}
K.~Hagiwara, K.~Hikasa and N.~Kai, Phys.\ Rev.\ D {\bf 27}, 84 (1983).
%
\bibitem{hhk1}
K.~Hagiwara, K.~Hikasa and N.~Kai, 
Phys.\ Rev.\ Lett.\ {\bf 52}, 1076 (1984);\ {\bf 67}, 931(E) (1991).
%
\bibitem{cs}
J.~C.~Collins and D.~E.~Soper,
Phys.\ Rev.\ D {\bf 16}, 2219 (1977).
%
\bibitem{csf}
We note that the Collins-Soper frame in Ref.~\cite{cdf} defines the 
 $y$-axis along $\vec{p}_p\times\vec{p}_{\bar{p}}$ and is 
opposite to our definition~\cite{hhk1}.  
This difference causes sign flips in $F_3$, $F_6$, $F_7$, and $F_8$.
%
\bibitem{chaichian}
M.~Chaichian, M.~Hayashi and K.~Yamagishi,
Phys.\ Rev.\ D {\bf 25}, 130 (1982);\ D {\bf 26}, 2534(E) (1982).
%
\bibitem{hky}
K.~Hagiwara, T.~Kuruma and Y.~Yamada, Nucl.\ Phys.\ B {\bf 369}, 171 (1992).
%
\bibitem{arnold2}
P.~B.~Arnold and R.~P.~Kauffman,
Nucl.\ Phys.\ B {\bf 349}, 381 (1991).
%
\bibitem{ellis1}
R.~K.~Ellis, D.~A.~Ross and S.~Veseli,
Nucl.\ Phys.\ B {\bf 503}, 309 (1997).
%
\bibitem{ellis2}
R.~K.~Ellis and S.~Veseli,
Nucl.\ Phys.\ B {\bf 511}, 649 (1998).
%
\bibitem{balas}
C.~Balas and C.~P.~Yuan, Phys.\ Rev.\ D {\bf 56}, 5558 (1997).
%
\bibitem{kulesza1}
A.~Kulesza and W.~J.~Stirling,
Eur.\ Phys.\ J.\ C {\bf 20}, 349 (2001).
%
\bibitem{kulesza2}
A.~Kulesza, G.~Sterman and W.~Vogelsang,
Phys.\ Rev.\ D {\bf 66}, 014011 (2002).
%
\bibitem{bawa}
A.~C.~Bawa and W.~J.~Stirling,
Phys.\ Lett.\ B {\bf 203}, 172 (1988).
%
\bibitem{kidonakis}
N.~Kidonakis and A.~S.~Vera, JHEP {\bf 0402}, 027 (2004).
%
\bibitem{mirkes2}
E.~Mirkes and J.~Ohnemus, Phys.\ Rev.\ D {\bf 50}, 5692 (1994).
%
\bibitem{boer}
A possibility of extending the soft-gluon resummation technique to the 
azimuthal angular distributions in small-$q_T$ region are recently 
investigated in  
D.~Boer and W.~Vogelsang, Phys.\ Rev.\ D {\bf 74}, 014004 (2006).
%
\bibitem{cteq}
J.~Pumplin {\it et al.}, JHEP {\bf 0207}, 012 (2002).
%
\bibitem{bases}
S.~Kawabata, Comput.\ Phys.\ Commun.\ {\bf 88}, 309 (1995).
%
\bibitem{cp}
Note that the azymuthal asymmetries such as Eqs.~(7) and (8) are
identical for $W^+$ and $W^-$ events, provided that we reverse the sign
of the pseudo-rapidity difference, $\Delta\eta$. Violation of this
rule signals $CP$ violation. 
See S.~Dawson and G.~Valencia, Phys.\ Rev.\ D {\bf 52}, 2717 (1995).
%
\end{thebibliography}
\end{document}